   \documentclass[final,1p,times]{elsarticle}
\usepackage{amssymb}
\usepackage{colortbl}
\usepackage[rightcaption]{sidecap}

\usepackage{nomencl}
\makenomenclature

\usepackage{amsmath}
\numberwithin{equation}{section}

\usepackage{soul}
\usepackage{subfigure}
\usepackage{graphicx}

\usepackage{enumerate}
\usepackage{multirow}
\usepackage{mathrsfs}
\biboptions{sort&compress}
\journal{Thermal Science and Engineering Progress}

\begin{document}

\begin{frontmatter}

\title{Instability of a plane Poiseuille flow bounded between inhomogeneous anisotropic porous layers}

\author[inst1]{Supriya Karmakar}

\affiliation[inst1]{organization={Department of Mathematics},%Department and Organization
            addressline={Indian Institute of Technology Madras}, 
            city={Chennai},
            postcode={600036}, 
            state={Tamil Nadu},
            country={India}}

\author[inst1]{Priyanka Shukla}

\begin{abstract}
%The linear stability analysis of a porous-fluid-porous plane Poiseuille flow with anisotropic and inhomogeneous permeability is performed. 

The linear stability analysis of a plane Poiseuille flow in a channel with anisotropic and inhomogeneous porous layers is performed. The generalized Darcy equation, along with the Navier–Stokes equation, governs the flow in the porous-fluid-porous channel. The Beaver--Joseph interface condition is assumed, which considers the coupling between the flows in the porous and fluid layers. The spectral collocation method is employed to solve the Orr--Sommerfeld type eigenvalue problem for the amplitude of the disturbances of arbitrary wavenumbers. The effect of anisotropy and inhomogeneous permeability on the stability characteristics is addressed in detail.
% 
%\st{It is found that decreasing the anisotropy and/or inhomogeneity of the porous layers destabilizes the flow. Therefore, the critical Reynolds number and associated critical wavenumber decrease and increase with decreasing anisotropy or inhomogeneity, respectively.}
% 
The stability characteristics of the anisotropy parameter (the ratio of permeability in the streamwise and the transverse direction) and the inhomogeneity function are presented in detail.
\end{abstract}

\begin{keyword}
interfacial flow \sep permeability \sep anisotropy \sep inhomogeneity \sep 
multi-layer flow
% %% PACS codes here, in the form: \PACS code \sep code
% %% MSC codes here, in the form: \MSC code \sep code
% %% or \MSC[2008] code \sep code (2000 is the default)
\end{keyword}

\end{frontmatter}



% \renewcommand{\nompreamble}{The next list describes several symbols that will be later used within the body of the document}
\nomenclature[1]{\(x,y\)}{rectangular coordinates}
\nomenclature[2]{\(j\)}{identifies a quantity associated with the $j^{th}$ porous layer}
\nomenclature[3]{\(h\)}{depth of open channel layer}
\nomenclature[4]{\(h_j\)}{depth of porous layer}
\nomenclature[5]{\(\mu\)}{viscosity}
\nomenclature[6]{\(\rho\)}{density}
\nomenclature[7]{\(\boldsymbol{u}\)}{velocity vector in the open channel layer}
\nomenclature[8]{\(\boldsymbol{u}_j\)}{volume averaged velocity vector in the porous layer}
\nomenclature{\(p\)}{pressure}
\nomenclature{\(p_j\)}{intrinsic average pressure}
\nomenclature{\(\epsilon_j\)}{porosity}
\nomenclature{\(\boldsymbol{K}\)}{permeability tensor}
\nomenclature{\(k_{jx}\)}{permeability in the $x$-direction}
\nomenclature{\(k_{jy}\)}{permeability in the $y$-direction}
\nomenclature{\(\eta_{jx}\)}{inhomogeneity function in the $x$-direction}
\nomenclature{\(\eta_{jy}\)}{inhomogeneity function in the $y$-direction}
\nomenclature{\( \xi_j\)}{anisotropy parameter}
\nomenclature{\(\mathcal{I}_j\)}{inhomogeneity factor}
\nomenclature{\(\alpha_j^{BJ}\)}{Beaver--Joseph coefficient at the porous-fluid interface}
\nomenclature{\(\sigma_j\)}{mean permeability}
\nomenclature{\(d_j\)}{thickness ratio between the $j^{th}$ porous layer and the open channel layer}
\printnomenclature
% ---------------------------------------------------------------------

\section{Introduction}
\label{sec:introduction}
Stability analysis of the flow through multi-layer porous fluid systems is a subject that has been studied extensively for the past many decades. Various applications, ranging from geophysical, industrial, technological, and biological systems employ multi-layer porous flow systems~\citep{Allen_1984collocation,blest_1999curing,nield2006convection,hussong2011continuum,ewing_1998numerical,Chen1988onset,chen1991onset,goharzadeh2005transition,berkowitz2002characterizing,MAKINDE2012chebyshev,DUKHAN2014,xiao_zhao_2013,majdalani2002two,chang1989velocity,voermans_ghisalberti_ivey_2017,kuwata_suga_2017,rosti_brandt_pinelli_2018}. For instance, the permeable coastal sediment flows, flows in fractured geological strata~\cite{goharzadeh2005transition,berkowitz2002characterizing,voermans_ghisalberti_ivey_2017,kuwata_suga_2017,rosti_brandt_pinelli_2018}, and gas flooding by injecting CO$_2$ or nitrogen to enhance oil recovery from oil reservoirs~\cite{Allen_1984collocation}. The manufacturing of cost-effective composite materials by resin film infusion (RFI), where resin flows, can be described by Darcy’s law. Such composite materials are used in aircraft or motor car production~\cite{blest_1999curing}. Another example of multi-layer porous-fluid systems is the blood flow through arteries, the gastrointestinal system, kidneys, and lungs~\cite{chang1989velocity,majdalani2002two,MAKINDE2012chebyshev}.
 For example, the stability analysis of blood flows in a large artery, by treating the large artery as a rigid channel with uniform width and the blood as an incompressible Newtonian fluid with variable viscosity due to transverse variation in hematocrit ratio, was performed by~\citet{MAKINDE2012chebyshev}. He observed that a transverse increase in the hematocrit ratio towards the arterial central region has a stabilizing effect.

\citet{chang2006instability}
performed the foremost stability work on a pressure-driven fluid flow overlying a saturated porous region. 
They numerically solved the Navier--Stokes--Darcy model and identified three different instability modes.~\citet{liu2008instability} solved the same problem using the Brinkman model and revealed two unstable modes in contrast to~\citet{chang2006instability}.~\citet{hill2008poiseuille} adopted a more generalized model by considering three regions in which a Darcy-Brinkman porous media is introduced between the
fluid and Darcy regions. The neutral stability curves are not found to be bimodal for specific parameter regimes.
% 
% ~\citet{silin_converti_dalponte_clausse_2011} examined this problem experimentally to validate the linear stability results.
% More recently, extending the work
% of~\citet{chang2006instability},~\citet{samanta2017role} incorporated a
% slippery wall instead of a fixed upper wall and studied the linear stability. His study revealed that the primary instability is started due to slip or depth ratio. Moreover, bimodal instability occurred due to the dominancy of porous and fluid modes.
% Detailed linear stability of porous-fluid-porous channel was analyzed by~\citet{tilton2006destabilizing,tilton2008linear} by considering the VANS equation~\cite{WhitakerB} along with the stress conditions involving interface coefficient given by~\citet{OCHOATAPIA1995a,OCHOATAPIA1995b}.
% 
% The main objectives of the present study are to extend the earlier study of~\citet{tilton2006destabilizing,tilton2008linear} by introducing anisotropy and inhomogeneity in the porous layers. {\color{red} ADD some more objective} 
Detailed linear stability of a plane Poiseuille flow in a confined porous channel was analyzed by~\citet{tilton2006destabilizing,tilton2008linear} by considering the volume-averaged Navier--Stokes equations (VANS) equation~\cite{WhitakerB} along with the shear stress conditions 
% involving interface coefficient 
given by~\citet{OCHOATAPIA1995a,OCHOATAPIA1995b}.~\citet{
tilton2006destabilizing,tilton2008linear}
% This authors 
suggested that inertial effects can be neglected if the ratio between the square root of permeability and channel half-height, i.e.~the mean permeability, is minimal. They concluded that the small increment in wall-permeability decreases the critical Reynolds number compared to the plane Poiseuille flow between rigid walls. Some other studies reported instability mechanism of porous-fluid configuration in the presence or absence of heat transfer (see,  e.g.,~\citet{straughan2008stability,nield2006convection}).

 In several practical situations and natural processes, a porous layer exhibits directional and spatial variation~\citep{straughan1996anisotropic,malashetty2010onset}. Therefore, it is deemed necessary to consider the effects of anisotropy and inhomogeneity while exploring the stability characteristics of a porous-fluid system.~\citet{deepu2015stability} 
 analyzed the impact of anisotropic and inhomogeneous permeability on the linear stability results proposed by~\citet{chang2006instability}. Their study reveals that a suitable wall-parallel or wall-normal permeability can control flow instability in porous channels.
However, choosing the optimal model for governing porous-fluid channels remains  unanswered~\cite{liu2008instability}.

 In this regard, the main objective of the present study is to investigate the effect of anisotropic and inhomogeneous permeability on the flow instabilities in a confined porous channel in isothermal conditions by using 
the generalized
Darcy’s equation along with the ~\citet{beavers1967boundary} boundary condition.
%is used in the present study.
%
The paper follows the following structure.  Section~\ref{sec:problem_formulation} describes the problem description and governing equations, the analytical expression of the base state solution, and the derivation of linear stability equations. 
The numerical method is explained in Sec.~\ref{sec:num_method}. 
The stability features for anisotropy and inhomogeneity are discussed in Sec.~\ref{sec:linear_stability_results}. Conclusions are in Sec.~\ref{sec:conclusion}.

%-------------------------------------------------------
\section{Mathematical formulation and governing equations}
\label{sec:problem_formulation}
\begin{figure}[!ht]
    \centering
	\includegraphics[scale=0.45]{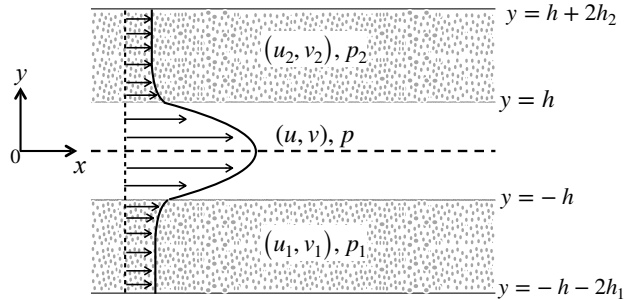}
	\caption{\small{Schematic of a two-dimensional plane Poiseuille flow in 
	a channel bounded between two porous layers.}}
	\label{fig:schematic}
\end{figure}

%--------------------------------------------------------
Consider a two-dimensional plane Poiseuille flow of an incompressible viscous fluid in a channel of height $2h$ bounded between two saturated porous layers of thickness
$2h_1$ and $2h_2$.
%, as shown in Fig.~\ref{fig:schematic}. 
A constant pressure gradient drives the flow. The lengths are measured from the center line of
the channel, as shown in figure~\ref{fig:schematic}.
The governing continuity and momentum balance equations in the fluid layer spanning $y\in (-h,h)$ read
\begin{subequations}
\begin{align}
	\nabla \cdot \boldsymbol{u}&=0,
	\label{eqn:channel_mass}\\
	\rho \left( \frac{\partial \boldsymbol{u}}{\partial t}+ \boldsymbol{u} \cdot \nabla \boldsymbol{u} \right)&=-\nabla p+\mu \nabla^2 \boldsymbol{u},
	\label{eqn:channel_NS}%
\end{align} 
\label{eqn:channel}%
\end{subequations}
where $t$ denotes the time, $x$ and $y$ are the streamwise and transverse directions, respectively, ${\boldsymbol{u}}=
(u, v)$,  $p$ and $\mu$ are  
the velocity vector, pressure and viscosity, respectively.

We consider highly dense porous layers, i.e., the permeability is low; hence the generalized Darcy’s law governs the momentum transport in the flow system~\citep{nield2006convection}.
The continuity and
generalized Darcy equation in the porous layers spanning $y \in \left(-h-2h_1, -h\right) \bigcup  \left(h, h+2h_2\right)$ 
read
\begin{subequations}
\begin{align}
	\nabla \cdot \boldsymbol{u}_j&=0,
	\label{eqn:porous_mass}\\
	\frac{\rho}{\epsilon_j}\frac{\partial \boldsymbol{u}_j}{\partial t}&=-\nabla p_j-\frac{\mu}{\boldsymbol{K}}\boldsymbol{u}_j,
	\label{eqn:porous_NS}%
\end{align} 
\label{eqn:porous}%
\end{subequations}
where $j=1$ and $2$ denotes the lower and upper layers, respectively, $\epsilon_j$ is the porosity, $\boldsymbol{u}_j=(u_j, v_j)$ is the volume averaged velocity vector, and $p_j$ is the intrinsic averaged pressure. The permeability tensor $\boldsymbol{K}$ reads
\begin{align}
\boldsymbol{K} =\begin{pmatrix}
k_{jx}\; \eta_{jx}\left({y}/{h} \right) & 0 \\
0   & k_{jy}\; \eta_{jy} \left({y}/{h}\right) \\
\end{pmatrix},
\end{align}
where $k_{jx}$ and $k_{jy}$ are the permeabilities in the $x$- and $y$-direction, respectively; $\eta_{jx}$ and $\eta_{jy}$ are the corresponding inhomogeneity functions depend only on $y$. The anisotropy parameter $\xi_j=k_{jx}/k_{jy}$ is defined as the ratio of the permeabilities. Note that the inhomogeneity functions $\eta_{jx}$, $\eta_{jy}$ can depend on $x$ and $y$ both; however, this gives rise to a two-dimensional base flow, thus making the problem complicated.
A linear inhomogeneity function is used by~\citet{green1969marginal}. However, in the present analysis, an inhomogeneous exponential  function~\cite{chen1992salt,deepu2015stability,deepu_dawande_basu_2015} of the following form is employed, 
\begin{align}
\eta_{jx}=\eta_{jy}=e^{\mathcal{I}_j\left( 1 \mp \frac{y}{h} \right)}, 
\label{eqn:inhomogeneityfunction}
\end{align}
where $\mathcal{I}_j$ is the inhomogeneity factor. Note that at lower and upper porous-fluid interfaces,  $\eta_{jx}(\mp h)=\eta_{jy}(\mp h)=1$. 
And the permeabilities in the $x$- and $y$-directions increase (decrease) with negative (positive) values of $\mathcal{I}_j$ for both porous layers.

No penetration conditions are applied at the bottom $y=-h - 2h_1$ ($j=1$) and top $y = h + 2h_2$ $(j=2)$ of the porous layers,
\begin{align}
	v_j=0.
	\label{eqn:nopenetration_bd}
\end{align}
%Al the interface, 
The continuity of velocity and balance of normal and shear stresses are imposed 
%at 
%the porous-fluid and fluid-porous interfaces, i.e.~
at the interfaces, which, respectively, read  
%and the momentum transfer condition at the interfaces $y=\pm h$,
\begin{equation*}
\refstepcounter{equation}
%\left.
\begin{aligned}
&v=v_j,\quad
p-2\mu \frac{\partial v}{\partial y}=p_j \quad\mbox{and}\quad
\frac{\partial u}{\partial y}=\pm \frac{\alpha_j^{BJ}}{\sqrt{k_{jx} \eta_{jx}(\mp h)}}\left(  u-u_j \right),
\end{aligned}
%\right\}
\eqno{(\theequation \textit{~a,b,c})}
\label{eqn:boundary_nd}%
\end{equation*}
%\begin{equation*}
%\refstepcounter{equation}
%\left.
%\begin{aligned}
%&v=v_j,\\
%&p-2\mu \frac{\partial v}{\partial y}=p_j,\\
%&
%\frac{\partial u}{\partial y}=\pm \frac{\alpha_j^{BJ}}{\sqrt{k_{jx} \eta_{jx}(\mp h)}}\left(  u-u_j \right),
%\end{aligned}
%\right\}
%\eqno{(\theequation \textit{a,b,c})}
%\label{eqn:boundary}
%\end{equation*}
where $\eta_{jx}(\mp h)=1$,  $\alpha_j^{BJ}$ denotes the Beaver--Joseph coefficient, which depends on the local structure of the porous materials near the fluid-porous interfaces~\citep{beavers1967boundary,goyeau2003momentum}. The positive and negative signs in~(\ref{eqn:boundary_nd}c) stand for the interface of lower and upper porous layers, respectively. 
%complete the problem. 
Note that~\citet{beavers1967boundary} condition~(\ref{eqn:boundary_nd}c) is employed, along with the streamwise component of permeability~\citep{chen1992salt}. 
%Eq.~\ref{eqn:boundary}(b,c) are continuity of normal stress, and normal velocity respectively.

%-------------------------------------------------------------------------
\subsection{Dimensionless governing equations and boundary conditions}
\label{subsec:dimensionless_problem}
% -------------------------------------------------------------------------
{

% The governing equations, along with the boundary conditions,~\eqref{eqn:channel}--\eqref{eqn:boundary_nd}, are normalized using the velocity, length and time scales as the mean channel velocity $U_m$, the half of the channel height $h$ and $h/U_m$, respectively. 
The following scaling is used for
%variables are made dimensionless as follows:
non-dimensionalization of the flow variables
\begin{align*}
    \left( x^{\star},y^{\star}\right)=\frac{1}{h}\left( x,y \right),\; t^{\star}=\frac{tU_m}{h}, \; \boldsymbol{u}^{\star}=\frac{\boldsymbol{u}}{U_m},\; p^{\star}=\frac{p}{\rho U_m^2},\; \boldsymbol{u}_j^{\star}=\frac{\boldsymbol{u}_j}{U_m},\;  p_j^{\star}=\frac{p_j}{\rho U_m^2},\; j=1,2,
\end{align*}
where $h$ is the half height of the open channel layer and $U_m$ is the mean channel velocity. After dropping $\star$, the non-dimensional form of~\eqref{eqn:channel}--\eqref{eqn:porous} is %read as
\begin{equation*}
\refstepcounter{equation}
		\frac{\partial \boldsymbol{u}}{\partial t} + \left(\boldsymbol{u} \cdot \nabla\right) \boldsymbol{u}= - \nabla p + \frac{1}{Re} \nabla^2 \boldsymbol{u}, \quad \nabla \cdot\boldsymbol{u} =0,
\eqno{(\theequation \textit{a,b})}
	\label{eqn:channel_nd}
\end{equation*}
\begin{equation*}
\refstepcounter{equation}
		\frac{\partial \boldsymbol{u}_j}{\partial t} = - \epsilon_j \nabla p_j + \frac{1}{\sigma_j^2 Re} \left( \frac{1}{\eta_{jx}(y)},\frac{\xi_j}{\eta_{jy}(y)}  \right) \odot \boldsymbol{u}_j, \quad \nabla \cdot\boldsymbol{u}_j =0,
\eqno{(\theequation \textit{a,b})}
	\label{eqn:porous_nd}
\end{equation*}
where $Re=\rho U_m h/\mu$ and $\sigma_j=\sqrt{k_{jx}}/h$ are the Reynolds number and the mean permeability, respectively, and $\xi_j=k_{jx}/k_{jy}$ is the anisotropy parameter. Here, $\odot$ represents the Hadamard product (or the element-wise product of two vectors), and the indices $j=1$ and $j=2$ refer to the lower $-1-2d_1<y<-1$ and the upper $1<y<1+2d_2$ porous layers, respectively. $d_j=h_j/h$ is the ratio of the thickness of the $j^{\rm th}$ porous layer to the open channel layer.
%
% The boundary conditions~\eqref{eqn:nopenetration_bd} and~\eqref{eqn:boundary_nd} can be rewritten in the dimensionless form as follows:
The non-dimensional interface conditions---continuity of velocity, normal and shear stresses conditions---at $y=\pm 1$ become
\begin{equation*}
\refstepcounter{equation}
		v=v_j,\quad
p-\frac{2}{Re} \frac{\partial v}{\partial y}=p_j \quad\mbox{and}\quad
\frac{\partial u}{\partial y}=\pm \frac{\alpha_j^{BJ}}{\sigma_j}\frac{1}{\sqrt{\eta_{jx}(\mp 1)}}\left(  u-u_j \right),
\eqno{(\theequation \textit{a,b,c})}
	\label{eqn:nd_interface}
\end{equation*}
% \begin{align}
%     &v=v_j,\quad
% p-\frac{2}{Re} \frac{\partial v}{\partial y}=p_j \quad\mbox{and}\quad
% \frac{\partial u}{\partial y}=\pm \frac{\alpha_j^{BJ}}{\sigma_j}\frac{1}{\sqrt{\eta_{jx}(\mp 1)}}\left(  u-u_j \right),
% \label{eqn:interface_nd}
% \end{align}
and no-penetration conditions at $y=\pm(1 + 2d_j)$ read as
\begin{align}
    v_j=0.
    \label{eqn:nd_nopenetration}
\end{align}
% The above formulations of using distinct governing equations to describe the dynamics of the porous and fluid channel regions demonstrates that the current investigation is based on a two-domain approach.

}
%% ------------------------------------------------------------
\subsection{Base flow}
\label{subsec:baseflow}
%% ------------------------------------------------------------
The undisturbed flow is assumed to be the steady $\left[ \partial \left( \cdot\right)/\partial t \right]$ and fully developed $\left[ \partial \left( \cdot\right)/\partial x \right]$ plane Poiseuille flow in both the channel and porous layers. 
% Under these assumptions solving non-dimensional form of~\eqref{eqn:channel}--\eqref{eqn:boundary}, 
Under these assumption solving the non-dimensionalized governing equations and boundary conditions~\eqref{eqn:channel_nd}--\eqref{eqn:nd_nopenetration},
we get the base state velocity $U_B=[U_2, U, U_1]$ and pressure $[P_2, P, P_1]$ in each layer as
\begin{equation*}
	\refstepcounter{equation}
%	\left.
	\begin{aligned}
{U}_2(y)&=-2A\sigma_2^2\eta_{2x}(y), \quad  1\leq y \leq 1+2d_2
		\\
{U}(y)&=Ay^2+By+C, \quad  -1\leq y \leq 1
		\\
{U}_1(y)&=-2A\sigma_1^2\eta_{1x}(y), \quad  -1-2d_1 \leq y \leq -1	
\\
P(x)&=P_1(x)=P_2(x) = \frac{1}{Re} \left( \frac{1}{1/6 + C/A}\right)x +\mbox{contant}	
	\end{aligned}
%	\right\}
\eqno{(\theequation \textit{~a,b,c,d})}	
\label{eqn:base_flow}
\end{equation*}
where 
% $Re=\rho U_m h/\mu$ is the Reynolds number,
% $\sigma_j=\sqrt{k_{ jx}}/h$ is the mean permeability, $d_j$ is the depth ratio $d_j=h_j/h$, and
%{\footnotesize
\begin{align}
\left.
\begin{aligned}
    \centering
%	&A=2\left( \frac{1}{6} - \frac{E}{G}  \right), \; B= \frac{A F}{G},\; C=\frac{A E}{G},\\
		&A=2\left( 1/6 - E/G  \right), \; B= AF/G,\; C=AE/G,
	 \\
	& E = \left( \frac{\alpha_1^{BJ} \left(\chi_1^2\sigma_1^2+1\right)}{\chi_1 \sigma_1}+1 \right) \left( \frac{\alpha_2^{BJ}}{\chi_2 \sigma_2} + 1 \right)+ \left( \frac{\alpha_2^{BJ} \left(\chi_2^2\sigma_2^2+1\right)}{\chi_2 \sigma_2}+1 \right) \left( \frac{\alpha_1^{BJ}}{\chi_1 \sigma_1} + 1 \right)  ,
	\\
	& F= \frac{\alpha_2^{BJ}}{\sigma_2 \chi_2} - \frac{\alpha_1^{BJ}}{\sigma_1 \chi_1}+\alpha_1^{BJ}\alpha_2^{BJ} \left( \frac{\chi_1 \sigma_1}{\chi_2 \sigma_2} - \frac{\chi_2 \sigma_2}{\chi_1 \sigma_1}\right) ,
	\\
	&G=\frac{\alpha_2^{BJ}}{\sigma_2 \chi_2}\left( 1+\frac{\alpha_1^{BJ}}{\sigma_1 \chi_1}  \right)+ \frac{\alpha_1^{BJ}}{\sigma_1 \chi_1}\left( 1+\frac{\alpha_2^{BJ}}{\sigma_2 \chi_2}  \right),
	\\
	&	\eta_{jx}=\eta_{jy}=e^{\mathcal{I}_j (1\mp y)},\\
	& \chi_1=\sqrt{\eta_{1x}(-1)}=1, \; \textnormal{and} \;\chi_2=\sqrt{\eta_{2x}(1)}=1.
	\end{aligned}
	\right\}
	\label{eqn:base_Coefficients}
\end{align}%
It is worth noticing that the base state velocity profile is independent of anisotropy parameter $\xi_j$. This is due to the fact that the $y$-momentum equation for the porous layer is identically satisfied under the base flow assumptions.

\begin{figure}[!ht]
    \centering
    \includegraphics[scale=0.22]{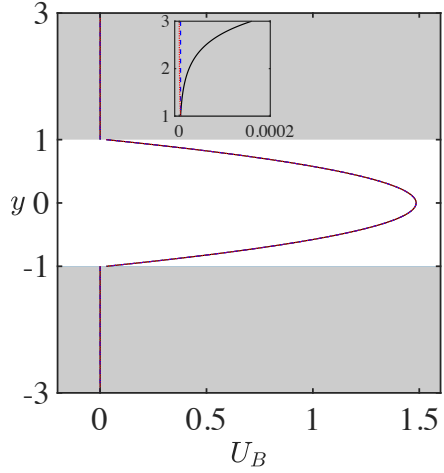}
    \caption{Base flow profile $U_{B}=[U_2,U,U_1]^{T}$ in both the open channel and porous layers for three different values of inhomogeneity factor: $\mathcal{I}_1=\mathcal{I}_2=-2$ (solid line), 0 (dashed line) and 2 (dotted line). The other parameters are fixed at $\sigma_1=\sigma_2=0.001,\; \epsilon_1=\epsilon_2=0.3,\; \alpha_1^{BJ}=\alpha_2^{BJ}=0.1$. The inset shows the zoomed part in the upper porous region.}
    \label{fig:base_flow}
\end{figure}

Figure~\ref{fig:base_flow} displays the base flow profile %for the confined porous channel %
%when 
of identical porous layers (i.e., the mean permeability, porosity, Beaver–Joseph coefficient and 
inhomogeneity factor are the same in both the porous layers)
%($\sigma_1=\sigma_2=0.001,\; \epsilon_1=\epsilon_2=0.3$ and $\alpha_1^{BJ}=\alpha_2^{BJ}=0.1, \mathcal{I}_1=\mathcal{I}_2$), %
%In the figure solid
%and 
for three different values of inhomogeneity factor
$\mathcal{I}_1=\mathcal{I}_2=\mathcal{I}=-2,0$ and $2$.  
Since all layers are of equal width $d_1=d_2=1$, the velocity profile is symmetric about the center-line of the channel. 
A jump in the interface velocity exists for all considered inhomogeneity factors due to the choice of interface condition~(\ref{eqn:boundary}c)
~\cite{beavers1967boundary}.
Note that for all values of inhomogeneity factor, the base flow profiles are parabolic and overlap in the open channel layer. However, the base velocity in the porous layers is uniform for the homogeneous porous layer ($\mathcal{I}=0$) but varies for the inhomogeneity of the porous layers ($\mathcal{I} \ne 0$). The base flow velocity profiles $U_j$ in porous layers depend on the inhomogeneity factor, whereas velocity $U$ of the fluid layer is independent of inhomogeneity ($\mathcal{I}$). This is due to the linear form of inhomogeneity function~\eqref{eqn:inhomogeneityfunction}, as it has a unit function value at the interface $y=\pm h$, i.e., $\chi_1=\chi_2=1$, see~\eqref{eqn:base_Coefficients}.
% 
%Also, we will see in subsection~\ref{subsec:linear_stability} that the base state velocity profile in the porous mediums (i.e., $U_1$ and $U_2$) do not explicitly involve in 
As shall see in the following subsection that the coupled linear stability equations do not depend on the base state velocity $U_j$ of the porous layers.

% -----------------------------------------------------
\subsection{Linear stability analysis}
\label{subsec:linear_stability}
% -----------------------------------------------------

The linear stability is examined by introducing an infinitesimal disturbance in the base flow ~\eqref{eqn:base_flow}.
%to perform the linear stability analysis.} 
Only 
two-dimensional disturbances are considered due to the Squires theorem, 
according to which 
the most unstable perturbations
are two-dimensional rather than three-dimensional~\citep{squire1933stability}. 
%It follows that 
%the two-dimensional perturbations can make a flow unstable at a lower Reynolds number $Re=\rho U_m h/\mu$ compared to the three-dimensional perturbations~\citep{squire1933stability}. 
A temporal linear stability analysis~\citep{schmid2002stability,drazin2004hydrodynamic} is performed by introducing  a normal-mode ansatz 
\begin{align}
%\left( \boldsymbol{u},\boldsymbol{u}_j,p,p_j  \right)= \left( U(y),U_j(y),p(x),p_j(x)\right)+ \left( \widehat{\boldsymbol{u}}(y),\widehat{\boldsymbol{u}}_j(y),\widehat{p}(y),\widehat{p}_j(y)  \right)e^{i\left( k x - k c t\right)},
\left(u, v, u_j, v_j, p, p_j  \right)= \left( U, 0, U_j, 0, P,P_j \right)+ \left( \widehat{u}, \widehat{v},
\widehat{u}_j,\widehat{v}_j, \widehat{p},\widehat{p}_j \right)e^{i\alpha \left(  x -  c t\right)},
\label{eqn:normal_mode}
\end{align}
where the hat variables are the amplitude of the disturbance; $\alpha$ is the real streamwise wavenumber, and $c$ is the complex phase-speed defined as $c=c_r+ic_i$. Note that a disturbance is unstable for $\alpha c_i>0$, stable for $\alpha c_i<0$, and neutrally stable for $\alpha c_i=0$. 
The disturbance equations are obtained by 
substituting~\eqref{eqn:normal_mode} into~\eqref{eqn:channel}--\eqref{eqn:boundary} and retaining only the first-order perturbation terms in the resulting equations.
Finally, the following set of 
linearized disturbance equations for $v$ and $v_j$ are achieved 
after eliminating pressure and horizontal velocity from the disturbance equations
%{\color{red} we get the following set of linearized disturbance equations for $v$ and $v_j$.}
\begin{subequations}
\begin{align}
&\left[  \left( {U}-c \right)\left( D^2-\alpha^2\right)-(D^2 {U})   \right]\widehat{v}-\frac{1}{i\alpha Re}\left( D^2-\alpha^2  \right)^2 \widehat{v}=0,
\label{eqn:stability_channel}
	\\[5pt]
&\frac{\epsilon_j}{i\alpha Re} \left[ \frac{D^2 \widehat{v}_j}{\eta_{jx}} - \xi_j \frac{\alpha^2 \widehat{v}_j}{\eta_{jy}} - \frac{D \eta_{jx} D \widehat{v}_j}{\eta_{jx}^2}\right]=c \sigma_j^2 \left( D^2 -\alpha^2\right) \widehat{v}_j, 
%	\\
%	\frac{\epsilon_1}{i\alpha Re} \left[ \frac{D^2 \widehat{v}_1}{\eta_{1x}} - \xi_1 \frac{\alpha^2 \widehat{v}_1}{\eta_{1y}} - \frac{D \eta_{1x} D \widehat{v}_1}{\eta_{1x}^2}\right]=c \sigma_1^2 \left( D^2 -k^2\right) \widehat{v}_1
\label{eqn:stability_porous}
\end{align}
\label{eqn:stability_eqn}
\end{subequations}
subject to the following boundary conditions
\begingroup
\allowdisplaybreaks
\begin{subequations}
\begin{align}
	&\widehat{v}_2=0  \quad &\hfill\textnormal{at}\quad\hfill y&=1+2d_2,
	\label{eqn:nopen_stability_lower}
	\\
	& \widehat{v}_2=\widehat{v} \quad  &\textnormal{at}\quad y&=1,
	\\
	& D^2 \widehat{v}=\frac{\alpha_2^{BJ} (D \widehat{v}-D\widehat{v}_2)}{\sigma_2 \sqrt{\eta_{2x}(1)}}  \quad  &\textnormal{at}\quad y&=1,
	\\
	& -\frac{D^3 \widehat{v}}{Re}+\left( \frac{3 \alpha^2}{Re}+i\alpha{U}  \right)D \widehat{v}-i\alpha(D{U})\widehat{v}-\frac{1}{\sigma_2^2Re}\frac{D\widehat{v}_2}{\eta_{2x}(1)}&
	\nonumber\\ & \qquad =i\alpha cD\widehat{v}-\frac{i\alpha c}{\epsilon_2}D\widehat{v}_2 \quad  &\textnormal{at}\quad y&=1,
	\\
	& -\frac{D^3}{Re} \widehat{v}+\left( \frac{3 \alpha^2}{Re}+i\alpha{U}  \right)D \widehat{v}-i\alpha(D{U})\widehat{v}-\frac{1}{\sigma_1^2Re}\frac{D\widehat{v}_1}{\eta_{1x}(-1)}&
	\nonumber\\ & \qquad=i\alpha cD\widehat{v}-\frac{i\alpha c}{\epsilon_1}D\widehat{v}_1 \quad  &\textnormal{at}\quad y&=-1,
	\\
	& D^2 \widehat{v}=-\frac{\alpha_1^{BJ}\left( D\widehat{v} - D \widehat{v}_1 \right)}{\sigma_1 \sqrt{\eta_{1x}(-1)}}  \quad  &\textnormal{at}\quad y&=-1,
	\\
	& \widehat{v}=\widehat{v}_1 \quad  &\textnormal{at}\quad y&=-1,
	\\
	&\widehat{v}_1=0 \quad  &\textnormal{at}\quad y&=-1-2d_1,
\label{eqn:nopen_stability_upper}
\end{align}
\label{eqn:boundary}
\end{subequations}
\endgroup
where $D=d/dy,\; D^2=d^2/dy^2, \; D^3=d^3/dy^3$; $\widehat{v},\; \widehat{v}_2,$ and $\widehat{v}_1$ are the amplitudes of the wall-normal velocity disturbances in the open channel layer, the upper porous and lower porous layers, respectively. 
%; $\widehat{v},\; \widehat{v}_1$, and $\widehat{v}_2$ are the amplitude of the wall-normal velocity disturbances in the channel layer, the lower, and the upper porous layers, respectively.
% Note that the base state velocity profile in the porous mediums (i.e., $U_1$ and $U_2$)  do not explicitly involve in the coupled disturbance equations.
%
% {\color{red}For simplicity, we assume 
% all layers of equal height (i.e.
% %the porous layers of equal height and same as open channel layer 
% $d_1=d_2=1$) along with the identical physical properties in porous layers
% ($\epsilon_1=\epsilon_2=\epsilon$, $\sigma_1=\sigma_2=\sigma$, $\alpha_1^{BJ}=\alpha_2^{BJ}=\alpha^{BJ}$, $\xi_1=\xi_2=\xi$, and $\mathcal{I}_1=\mathcal{I}_2=\mathcal{I}$). The flow is governed by the dimensionless parameters, namely, $Re,\; \sigma,\; \epsilon,\;\alpha^{BJ},\; \xi$ and $\mathcal{I}$.}
{For simplicity, we assume porous walls of identical heights and physical properties, i.e.~$d_1=d_2=d$, $\epsilon_1=\epsilon_2=\epsilon$, $\sigma_1=\sigma_2=\sigma$, $\alpha_1^{BJ}=\alpha_2^{BJ}=\alpha^{BJ}$, $\xi_1=\xi_2=\xi$ and $\mathcal{I}_1=\mathcal{I}_2=\mathcal{I}$. The flow is controlled by the dimensionless parameters, namely, $Re,\;d,\; \sigma,\; \epsilon,\;\alpha^{BJ},\; \xi$ and $\mathcal{I}$.}

%-------------------------
\section{Numerical method}
\label{sec:num_method}
%%--------------------------------------------
The coupled linear stability problem~\eqref{eqn:stability_eqn} along with the boundary conditions~\eqref{eqn:nopen_stability_lower}-\eqref{eqn:nopen_stability_upper} is solved numerically using the Chebyshev spectral collocation method~\citep{schmid2002stability,Canuto_1988}.
We approximate the disturbance amplitudes $\widehat{v}_2,\; \widehat{v}$ and $\widehat{v}_1$ using the Chebyshev expansions in terms of $N$ Chebyshev polynomials as
\begin{align}
	\widehat{v}_2=\sum_{n=0}^{N} a_n T_n\left( y^{\prime}\right),\; \widehat{v}=\sum_{n=0}^{N} b_n T_n\left( y^{\prime}\right) \; \textnormal{and} \; \widehat{v}_1=\sum_{n=0}^{N} c_n T_n\left( y^{\prime}\right),
	\label{eqn:cheb_expansion}
\end{align}
where $a_n,\; b_n$ and $c_n$ are the unknown, called Chebyshev coefficients, to be determined, and $T_n\left( y^{\prime}  \right)$ is the $n$-th Chebyshev polynomial defined as
\begin{align}
	T_n\left( y^{\prime}  \right)=\cos \left(  n \theta \right),\;\theta=\cos^{-1} \left( y^{\prime} \right),\quad -1 \leq y^{\prime} \leq 1.
\end{align}
As the Chebyshev polynomials $T_n(y^{\prime})$ are defined over the domain $\left[ -1,1\right]$, the following transformation is employed to transform the physical domain $y\in[-3,3]$ to the computational domain $y^{\prime}\in \left[ -1,1\right]$
\begin{align}
	y^{\prime}=\begin{cases} 
      \left(y- d_2-1\right)/{d_2} & \; 1 \leq y\leq 1+2d_2 \\
      y   & -1 \leq y \leq 1 \\
      \left(y+ d_1+1\right)/{d_1}  & -1-2d_1\leq y\leq -1.
      \end{cases}
    \label{eqn:spectral_mapping}
\end{align}
The expansions~\eqref{eqn:cheb_expansion} are substituted into~\eqref{eqn:stability_eqn}, and evaluated at the Gauss-Lobatto points $y_j=\cos\left(j\pi/N \right)$, which are extrema of the Chebyshev polynomials, where $j=0,1,2,\cdots,N$. The discretization of~ \eqref{eqn:stability_eqn}--\eqref{eqn:boundary} leads to a generalized eigenvalue problem
\begin{align}
    \mathbf{A}{X}=\omega \mathbf{B}{X},
    \label{eqn:combined_matrix_discretized}
\end{align}
where $\omega=\alpha c$ is the eigenvalue and $X$ is the associated eigenvector.  Equation~\eqref{eqn:combined_matrix_discretized} represents $3N+3$ equations for the $3N+3$ unknown Chebyshev coefficients $a_j, b_j, c_j$. To impose the boundary conditions, we follow~\citet{schmid2002stability} and substitute the Chebyshev expansions~\eqref{eqn:cheb_expansion} into~\eqref{eqn:boundary}. The eight boundary conditions are imposed by replacing the corresponding eight rows of $\mathbf{A}$ and $\mathbf{B}$ matrices of~\eqref{eqn:combined_matrix_discretized}.

% These discretized block matrices $\mathbf{A}$ and $\mathbf{B}$ are shown in~\ref{appdx:stability_mat}.

The corresponding disturbance amplitudes $\widehat{\boldsymbol{v}}=\left[ \widehat{v}_2, \widehat{v}, \widehat{v}_1 \right]^{T}$ is found using the Chebyshev coefficients and normalized by using the maximum fluid layer velocity $\left( \max_{y \in \left[ -1,1 \right]} \big| \widehat{v}\big| \right)$.
% \begin{align}
%     \max_{y \in \left[ -1,1 \right]} \big| \widehat{v}\big|=\widehat{v}_{max}.
% \end{align}
 To check the convergence of the spectrum obtained numerically, a relative error is used as defined below
 %in~\citet{tilton2008linear,supriyaStability_2022}
\begin{align}
	e_{N}=\frac{\|\boldsymbol{\Omega}_{N+1}-\boldsymbol{\Omega}_{N}\|_{2}}{\| \boldsymbol{\Omega}_{N}\|_{2}},
\end{align}
where $\|\cdot \|_{2}$ represents the $L_2$ norm, and $\boldsymbol{\Omega}_{N}$ and $\boldsymbol{\Omega}_{N+1}$ are vectors containing twenty least stable eigenvalues obtained using $N$ and $N+1$ collocation points.

\begin{figure}[!ht]
	\centering
	\includegraphics[scale=0.2]{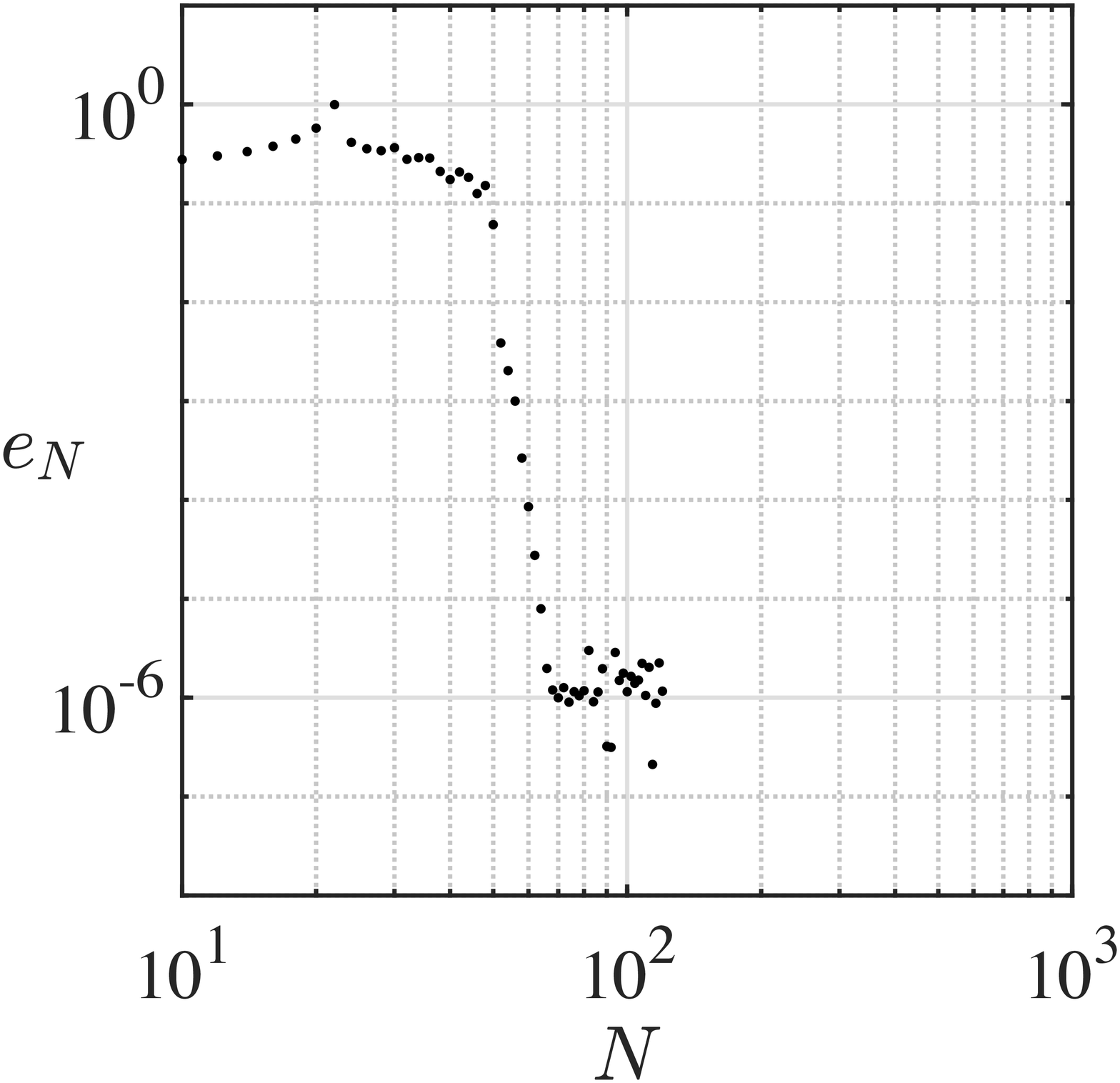}
	\caption{The variation of relative error $e_{N}$ with the number $N$ of Chebyshev polynomial in the $\log$ scale, when $Re=2000,\;d=1,\; \alpha=1,\; \sigma=0.001,\; \epsilon=0.3,\; \alpha^{BJ}=0.1,\; \xi=1$ and $\mathcal{I}=0$.}
	\label{fig:collocation_points}
\end{figure}
Figure~\ref{fig:collocation_points} displays the variation of relative error $e_{N}$ for isotropic $\xi=1$ and homogeneous $\mathcal{I}=0$ porous layers as a function of collocation points $N$. 
It is seen that $e_{N}$ saturates to $O(10^{-6})$ for $N\ge 70$. For other values of $\xi$ and $\mathcal{I}$, similar behavior of $e_{N}$ is 
observed (figures not presented).
In addition, at 
%this value of mean permeability 
$\sigma=0.001$, the thickness ratio
%of the thickness of the porous layer to the fluid layer 
$d$ has a negligible effect on the %needed number of 
collocation points $N$.
Therefore, $N \approx 80$ is adequate to obtain numerical accuracy at $\sigma=0.001$. 
% {\color{red} Comment on the variation of the colocation points for different depth ratios.}

%------------------------------------------------------------
\section{Results and Discussion}
\label{sec:linear_stability_results}
%------------------------------------------------------------

\begin{figure}[!htbp]
    \centering
     \includegraphics[scale=0.3]{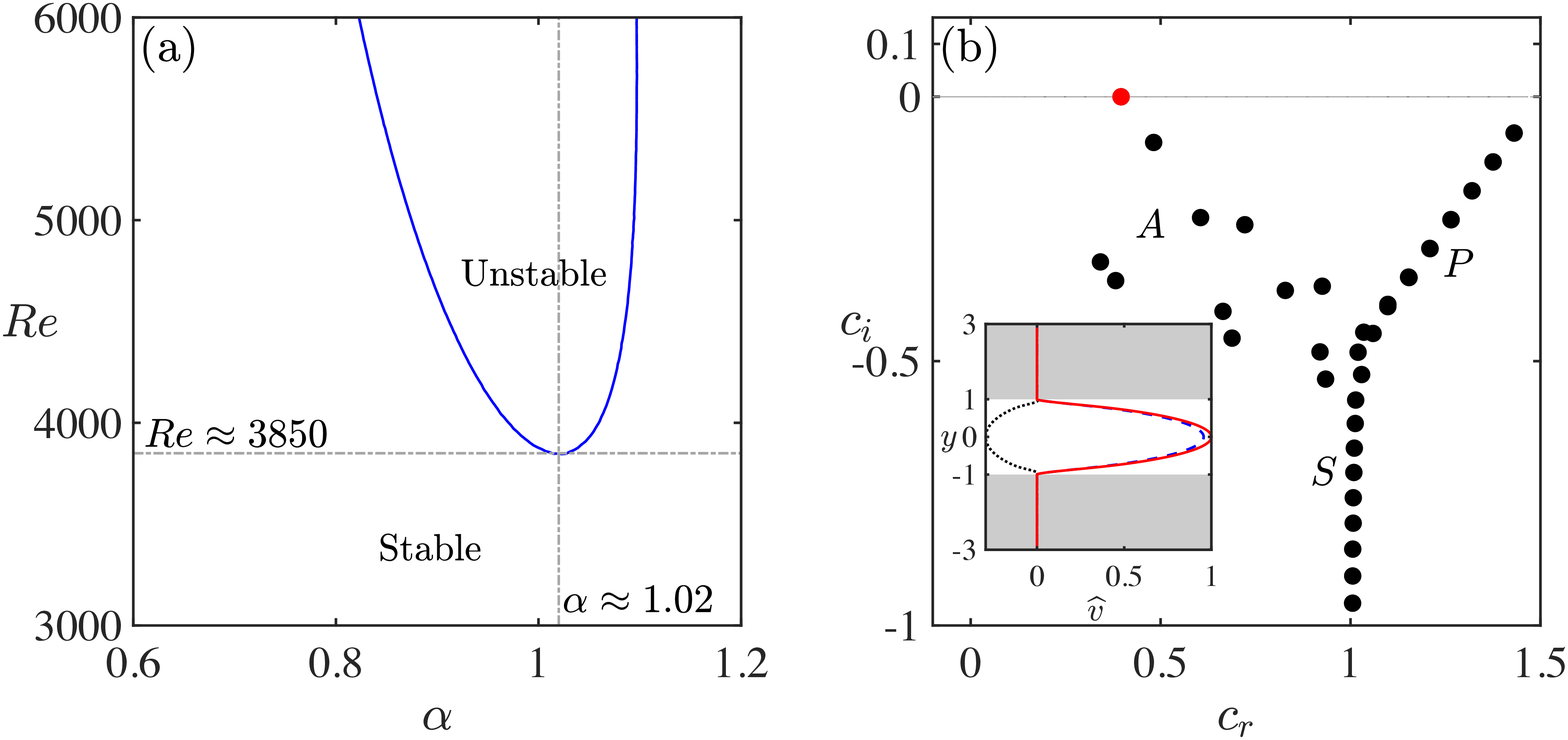}
    \caption{\small{\textbf{(a)} The neutral stability curve in the $(\alpha, Re)$--plane for $\sigma=10^{-5}$; \textbf{(b)} Spectrum at the critical parameters $Re_c=3850,\; \alpha_c=1.02$. The inset shows the eigenfunction, $| \widehat{\boldsymbol{v}}|$ (Solid line), $\Re{\left(\widehat{\boldsymbol{v}}\right)}$ (dotted line), $\Im{\left(\widehat{\boldsymbol{v}}\right)}$ (dashed line), of the marked eigenvalue. The other parameters are fixed at $\epsilon=0.3,\; \alpha^{BJ}=0.1,\; \xi=1,\; \mathcal{I}=0$.}}
    \label{fig:poiseuille}
\end{figure}

%The whole analysis has carried out by considering the fixed value of porosity $\epsilon=0.3$, mean permeability $\sigma=0.001$ and Beaver-Joseph coefficient $\alpha^{BJ}=0.1$, as used in previous studies~\cite{chang_chen_straughan_2006,chang2006thermal,liu2008instability}. 
The present numerical results are tested for validation with the available result of the classical plane Poiseuille flow~\cite{orszag_1971}. 
Figure~\ref{fig:poiseuille}(a) depicts the neutral stability curve $(\omega_i=0)$ in the $\left(\alpha,Re\right)$-plane for $\sigma=10^{-5}$, with other parameters being set as $\epsilon=0.3,\; \alpha^{BJ}=0.1,\; \xi=1$ and $\mathcal{I}=0$. This set of parameters recovers stability results for classical plane Poiseuille flow. The flow is unstable $(\omega_i>0)$ inside the neutral stability curve and stable $(\omega_i<0)$ outside. 
The critical Reynolds number $Re_c$ is found to be 3850, which occurs at wavenumber $\alpha_c=1.02$.
Note that the critical Reynolds number $Re_c$ is defined as the lowest Reynolds number at which instability is triggered. Figure~\ref{fig:poiseuille}(b) depicts the eigenvalue and the eigenfunctions (inset) of the most unstable eigenvalue (marked with the red dot in the main panel) at the critical parameters $(Re_c,\alpha_c)=(3850,1.02)$. 
The spectrum also shows the $A,\; P$ and $S$ branches of the classical plane Poiseuille flow~\cite{schmid2002stability}. 
It is to be noted from the eigenfunction (see the inset) of the most unstable eigenvalue that the amplitude of the perturbation velocity vanishes at the interfaces and the porous regions; thus, we recover the TS wave observed in a channel flow with impermeable walls. These results show an excellent agreement between present results with mean permeability $\sigma \approx 0$ and the classical plane Poiseuille flow. 

Following previous studies~\cite{chang_chen_straughan_2006,chang2006thermal,liu2008instability}, we now fix porosity, mean permeability, and Beaver-Joseph coefficient to $\epsilon=0.3$, $\sigma=0.001$, and $\alpha^{BJ}=0.1$, respectively.
%in all results presented here. 
In Secs.~\ref{subsec:anisotropy} and~\ref{subsec:inhomogeneity}, we analyze the effect of anisotropy $\xi$ and inhomogeneity $\mathcal{I}$ with respect to the depth ratio $d$, respectively.

\subsection{Effect of anisotropy}
\label{subsec:anisotropy}
% -------------------------------------------
% ------------------------------------------------------

% ----------------------------------------------------------
{This subsection studies the effect of anisotropy $\xi$ with respect to the depth ratio $d$ %by considering 
when the porous layers %to be 
are homogeneous (i.e.,~$\eta_x=\eta_y=1$).
\begin{figure}[!htbp]
    \centering
    \includegraphics[scale=0.3]{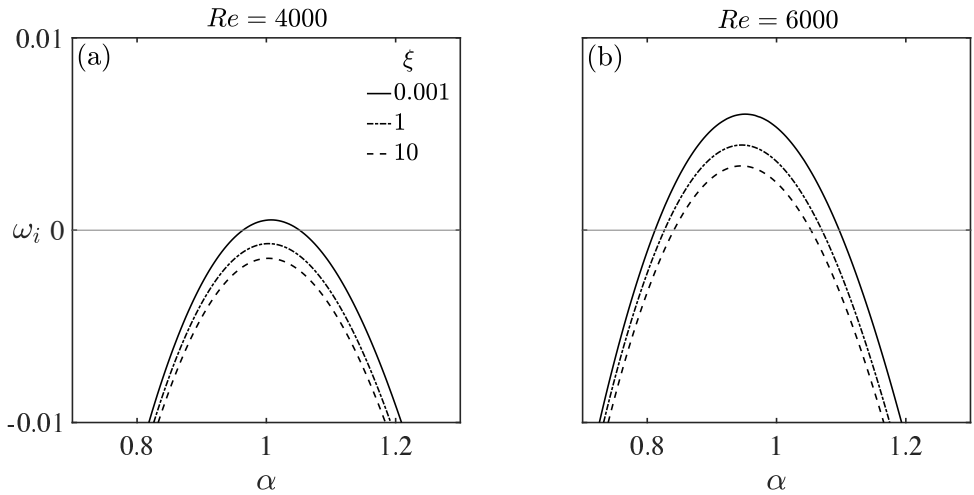}
    \caption{Variation of the growth rate $\omega_i$ with the wavenumber for $\xi= 0.001$ (solid line), $1$ (dash-dotted line) and $10$ (dashed line) at $d=1$. Left and right panels represent results for (a) $Re=4000$, and (b) $Re=6000$.}
    \label{fig:dispersion_anisotropy}
\end{figure}
Figure~\ref{fig:dispersion_anisotropy} displays the anisotropy effects on the growth rate $\omega_i$ of the most unstable mode when $d=1$ for two values of Reynolds number $Re=4000$ [panel (a)] and $Re=6000$ [panel (b)]. For both these cases, the growth rates decrease as $\xi$ increases for a fixed wavenumber $\alpha$. The configuration remains stable when $Re=4000$ and $\xi \ge 1$  [panel a].
%figure~\ref{fig:dispersion_anisotropy}(a)]. 
Moreover, at this $Re$, when the relative magnitude of the cross-stream permeability is larger (i.e., $\xi=0.001$), the growth rates become positive at wavenumber $\alpha \approx O(1)$, and the system becomes unstable. As soon as the Reynolds number $Re$ increases to $6000$, the growth rates gradually increase and become positive at each $\xi$ [panel b].
%[see figure~\ref{fig:dispersion_anisotropy}(b)]. 
%The system destabilizes in a bandwidth of unstable wavenumbers for various $\xi$ [see panel (a) at $\xi=0.001$ and panel (b) at each $\xi$ of figure~\ref{fig:dispersion_anisotropy}]. 
Thus, depending on the anisotropy parameter and Reynolds number, the system  destabilizes in a bandwidth of unstable wavenumbers.   
It is also seen from figure~\ref{fig:dispersion_anisotropy} that at fixed $Re$, 
%when compared to both isotropic ($\xi=1$) and anisotropic ($\xi \neq 1$) porous walls, 
decreasing and increasing the anisotropy parameter $\xi$ enhances and suppresses the growth rate of the most unstable mode and, therefore, increases and decreases the bandwidth of unstable wavenumbers.

\begin{figure}[!htbp]
    \centering
    \includegraphics[scale=0.31]{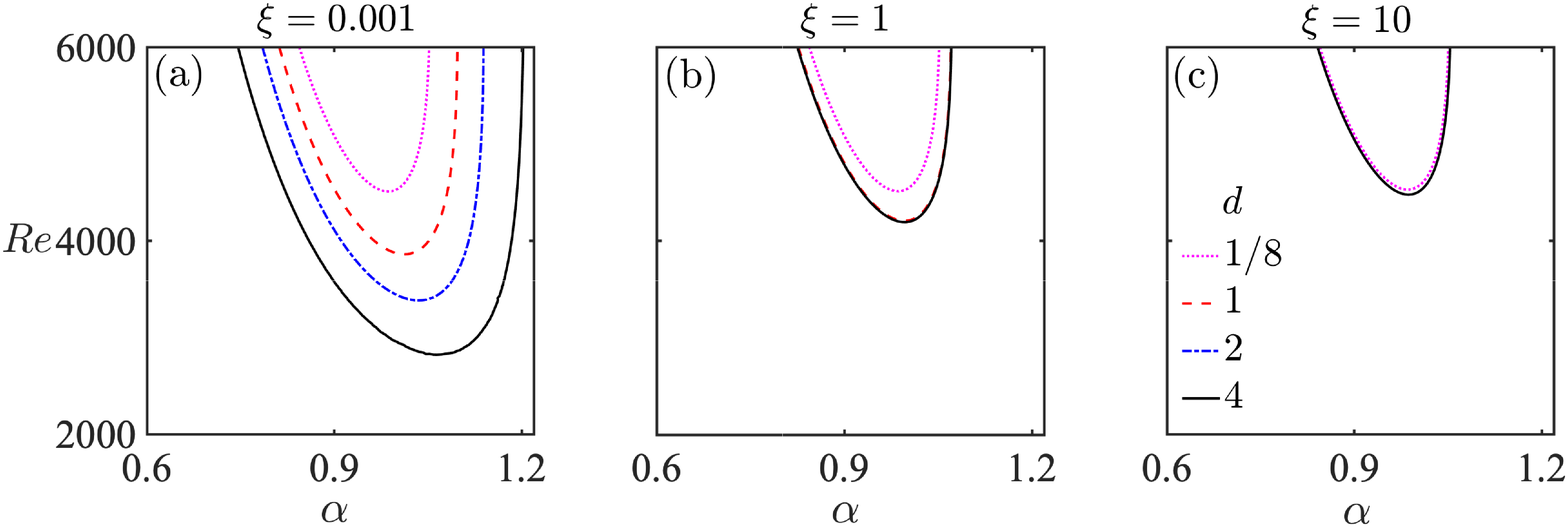}
    \caption{Neutral stability curves in $(\alpha,Re)$-plane for various depth ratios $d$ with $\xi$ as (a) $0.001$, (b) $1$, and (c) $10$.}
    \label{fig:neutral_anisotropy}
\end{figure}
To examine the effect of depth ratio $d$, figure~\ref{fig:neutral_anisotropy} shows the neutral stability curves 
for four values of $d$, with the
anisotropy parameter being fixed at 
$\xi= 0.001$ [panel a], $1$ [panel b], $10$ [panel c].  
%and various depth ratios $d$. 
Note that the unstable regions become larger and the range of unstable wavenumbers increases as $d$ increases at any fixed $\xi$. When $\xi \ge 1$, the neutral stability curves almost coincide, and thereby the corresponding critical Reynolds numbers are practically insensitive to the depth ratio [panels b and c]. Furthermore, at each fixed depth ratio $d$, the unstable region widens when the relative magnitude of the streamwise permeability is small [panel a] relative to the cross-stream permeability as against the case when the relative magnitude of the streamwise permeability are larger [panel c] relative to the cross-stream permeability. It is also evident from figure~\ref{fig:neutral_anisotropy} that a single mode characterizes the instability domain for all
$\xi$ and $d$ considered here. Therefore, the neutral stability curves are unimodal, similar to the classical plane Poiseuille flow~\cite{schmid2002stability}.

\begin{table}[ht]
\centering
 \begin{tabular}{ |c||p{1.4 cm}|p{1.4 cm}|p{1.4 cm}|p{1.4 cm}|p{1.4 cm}|p{1.4 cm}|}
 \hline
 \multirow{2}{4em}{$\qquad d$}&\multicolumn{2}{|c|}{$\xi=0.001$}& \multicolumn{2}{|c|}{$\xi=1$} &\multicolumn{2}{|c|}{$\xi=10$}\\
 \cline{2-7}
         & $Re_c$& $\alpha_c$& $Re_c$& $\alpha_c$& $Re_c$ & $\alpha_c$\\
 \hline \hline
1/8 & 4509.38 & 0.9864 & 4511.71 & 0.9864 & 4528.78 & 0.9862\\
1   & 3860.96 & 1.0109 & 4205.73 & 0.9950 & 4476.48 & 0.9870\\
2   & 3384.05 & 1.0340 & 4192.15 & 0.9960 & 4476.48 & 0.9860\\
4   & 2816.99 & 1.0570 & 4191.89 & 0.9960 & 4476.43 & 0.9860\\
\hline
\end{tabular}
\caption{The effect of anisotropy parameter $\xi$ and depth ratio $d$ on the critical Reynolds number and critical wavenumber.}
\label{table:neutral_s_d}
\end{table}

For more understanding of criticality, 
the critical Reynolds number $Re_c$ and critical wavenumber $\alpha_c$ are obtained for various $\xi$, and $d$ values, and illustrated in Table~\ref{table:neutral_s_d}. 
% From the table, it is seen that there is a $0.052\%$ ($0.378\%$) reduction (increment) in $Re_c$ at $\xi=0.001\; (\xi=10)$ as compared to its value at $\xi=1$ at small depth ratio $d=1/8$.
% From the table, it is seen that there is a $0.052\%$ reduction in $Re_c$ at $\xi=0.001$ as compared to its value at $\xi=1$ at small depth ratio $d=1/8$.
The table indicates that $Re_c$ reduces by $0.052\%$ at $\xi=0.001$ compared to its value at $\xi=1$ for a small depth ratio $d=1/8$. The same trend is observed at higher $d=4$ (reduces $32.799\%$ for $\xi=0.001$ compared to $\xi=1$) when the thickness of porous layers is significantly greater than the fluid layer. However, when the relative magnitude of the streamwise permeability is greater than the relative magnitude of the cross-stream permeability, i.e., when $\xi>1$, $Re_c$ increases compared to the isotropic case ($\xi=1$) for all values of depth ratio $d$. Particularly, for $\xi=10$, there is an increase of $0.378\%$ and $6.788\%$ at $d=1/8$ and $d=4$, respectively.

Observe that the anisotropy parameter $\xi$ and the depth ratio $d$ offer a viable control technique for instability in a confined porous--fluid--porous channel. To suppress (enhance) the instability in such a system, one must design porous walls of various widths with the cross-stream permeability being smaller (more) than the streamwise permeability.
}

\begin{figure}[!ht]
    \centering
    \includegraphics[scale=0.29]{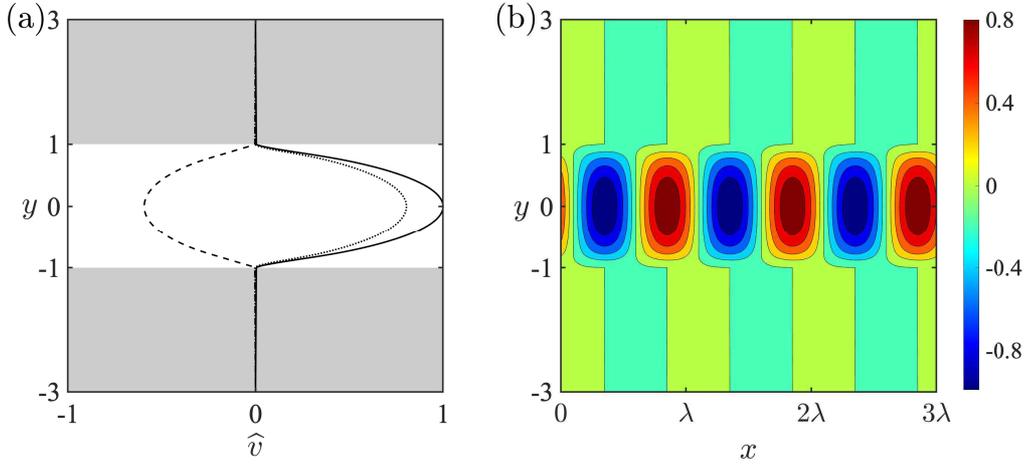}
    \caption{(a) The normalized eigenfunction,  $| \widehat{\boldsymbol{v}}|$ (Solid line), $\Re{\left(\widehat{\boldsymbol{v}}\right)}$ (dotted line), $\Im{\left(\widehat{\boldsymbol{v}}\right)}$ (dashed line), when $\xi=0.001,\; Re_c=3860.96$ and $\alpha_c=1.0109$.  (b) The contour plot of associated stream function in the $(x,y)$-plane. Here $\lambda=2\pi/\alpha_c$.}
    \label{fig:eigenfunc_anisotropy}
\end{figure}

{ It is crucial to study the dominant mode (mode with maximum growth over all wavenumbers) of instability.}
%, which may be captured in controlled experiments. }
%further depth the single mode of instability where the primary instability occurs. 
In this regard, the dominant instability mode at $(\alpha_c,Re_c)=(1.0109,3860.96)$ is chosen for $\xi=0.001$ and $d=1$. 
For these parameters, the normalized eigenfunction [depicted in figure~\ref{fig:eigenfunc_anisotropy}(a)] displays the significant variation in the open channel layer; therefore, this dominant mode is a fluid-layer mode. The corresponding contour of stream function perturbation in the $(x,y)$-plane is shown in figure~\ref{fig:eigenfunc_anisotropy}(b). Here $x\in [0,3\lambda]$ and $y\in [-1-2d,1+2d]$, where $\lambda$ represents the unit wavelength. Note that the variation of normalized eigenfunction generates a series of
identically shaped vortices at the center of the open channel layer, which may be responsible for the primary instability. This fact fully supports the priority of the dominant fluid-layer mode on the primary instability, as predicted in figure~\ref{fig:neutral_anisotropy}. Similarly, for the other values of $\xi$ and $d$, the dominant mode of instability has an enormous variation in the open channel-layer.

% ----------------------------------------------------------------------------

\begin{figure}[!htbp]
    \centering
    \includegraphics[scale=0.3]{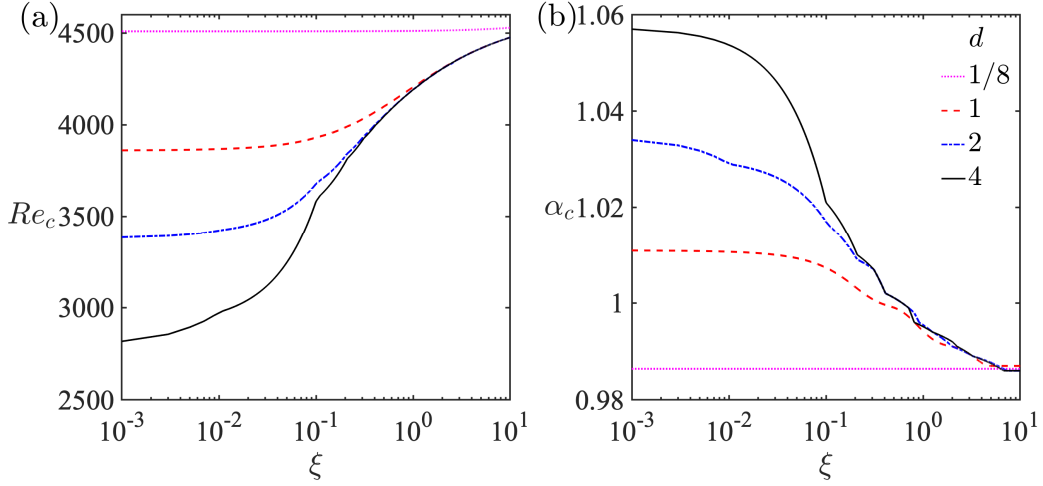}
    \caption{Variation of the (a) critical Reynolds number and (b) critical wavenumber with the anisotropy parameter $\xi$ for various depth ratio $d$.}
    \label{fig:critical_anisotropy}
\end{figure}

{ To summarize effect of anisotropy and depth ratio on the onset of instability, the variation of the critical Reynolds number $Re_c$ and critical wavenumber $\alpha_c$ with the anisotropy parameter $\xi$ for various depth ratio $d$ are displayed in figures~\ref{fig:critical_anisotropy}(a) and~\ref{fig:critical_anisotropy}(b), respectively. While $Re_c$ increases as $\xi$ increases for all depth ratios considered here, the critical wavenumber $\alpha_c$ decreases with the increase in $\xi$. It is also seen from figure~\ref{fig:critical_anisotropy}(a) that at a fixed depth ratio $d$, there is a cutoff value $\xi_{m}$ of anisotropy parameter below which $Re_c$ remains unchanged and above which it increases rapidly with increasing anisotropy parameter $\xi$; therefore, the neutral stability curves for various $\xi \le \xi_m$ values merge (figure not shown). Thus, the destabilizing effects of anisotropy parameter $\xi$ are insignificant with further reduction in $\xi$ from $\xi_m$.
Recall that a reduction in the anisotropy parameter $\xi$ is associated with either a relative increase in the permeability in the wall-normal direction ($k_y$) or a relative decrease in the permeability in the wall-parallel direction ($k_x$), which reduces the flow resistance, enhances the disturbance growth rate, and destabilizes the flow. The above observations reveal the significant destabilization of the system with anisotropic walls relative to the channel with porous walls as the relative magnitude of cross-stream permeability increases. Therefore, anisotropic permeability relative to isotropic permeability provides a potential strategy to control flow instability. 
In addition, the critical Reynolds number $Re_c$ (the critical wavenumber $\alpha_c$) decreases (increases) for each fixed $\xi$ as $d$ increases. As the porous layer thickness increases (decreases) relative to the fluid layer thickness, the system becomes unstable (stable), and this is true for the Poiseuille flow in a channel with either isotropic or anisotropic walls.
}

\subsection{Effect of inhomogeneity}
\label{subsec:inhomogeneity}
% -------------------------------------------

This subsection studies the effect of inhomogeneity $\mathcal{I}$ concerning the depth ratio $d$ while keeping the anisotropy parameter $\xi$ fixed. 
It is worth noticing that changing $\mathcal{I}$ is equivalent to changing the overall permeability of the porous medium. It can be seen from~\eqref{eqn:stability_porous} that in contrast to the anisotropy, the effect of inhomogeneity on the stability of the flow is more unpredictable since it generally appears in a complex way, as shown in this section. 
%This is because of the form of inhomogeneity function, which appears
{
\begin{figure}[!htbp]
    \centering
    \includegraphics[scale=0.3]{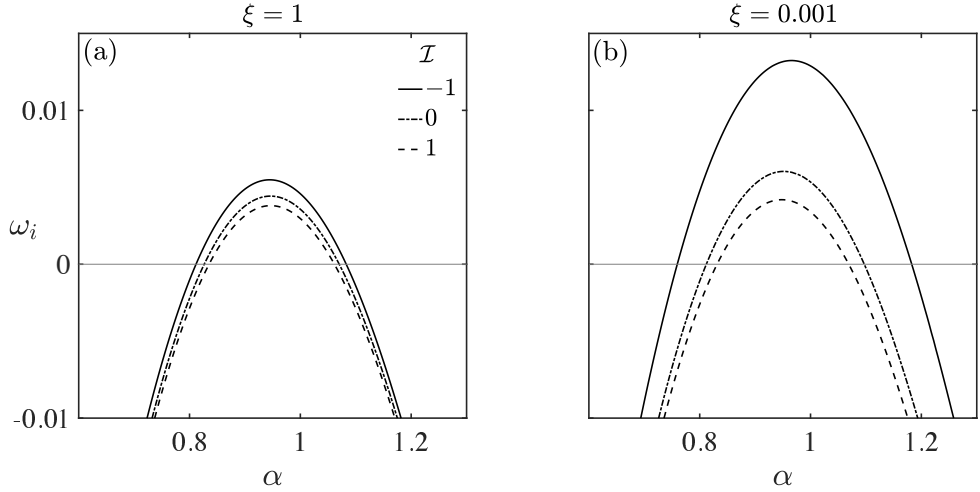}
    \caption{The temporal growth rate $\omega_i$ as a function of the wavenumber $\alpha$ for various values of inhomogeneity parameter $\mathcal{I}$; (a) $\xi=1$ and (b) $\xi=0.001$. The other parameters are fixed as $Re=6000,\; d=1$.}
    \label{fig:dispersion_inhomogeneity}
\end{figure}

Figure~\ref{fig:dispersion_inhomogeneity} shows the effect of the inhomogeneity factor on the growth rate of the most unstable mode when $d=1$ for two values of the anisotropy parameter $\xi=1$ [panel a] and $\xi=0.001$ [panel b]. It is seen from figure~\ref{fig:dispersion_inhomogeneity}(a) that the growth rates enhance as $\mathcal{I}$ decrease for a fixed wavenumber $\alpha$ and thereby destabilizing the flow. This destabilization effect is striking for $\xi<1$ when inhomogeneity factor decreases, see figure~\ref{fig:dispersion_inhomogeneity}(b).
% Moreover, as the anisotropy parameter $\xi$ decreases from $\xi=1$ to $\xi=0.001$, the growth rate further enhances for a fixed value of inhomogeneity factor $\mathcal{I}$ and wavenumber $\alpha$. 
Therefore, for any fixed $\xi$ when compared to both homogeneous ($\mathcal{I}=0$) and inhomogeneous ($\mathcal{I}\neq 0$) porous walls, decreasing and increasing the inhomogeneity factor $\mathcal{I}$ enhances and suppresses the growth rate of the most unstable mode, which increases and decreases the bandwidth of unstable wavenumbers.

% The stabilizing effects of anisotropic permeability is shown in the above section~\ref{subsec:anisotropy}, here in this subsection
Since the linear stability results of anisotropy parameter $\xi$ are shown in section~\ref{subsec:anisotropy}, we thus focus on the effect of inhomogeneity factor $\mathcal{I}$ when $\xi=1$ to recover inhomogeneous but isotropic porous medium.
\begin{figure}[!htbp]
    \centering
    \includegraphics[scale=0.3]{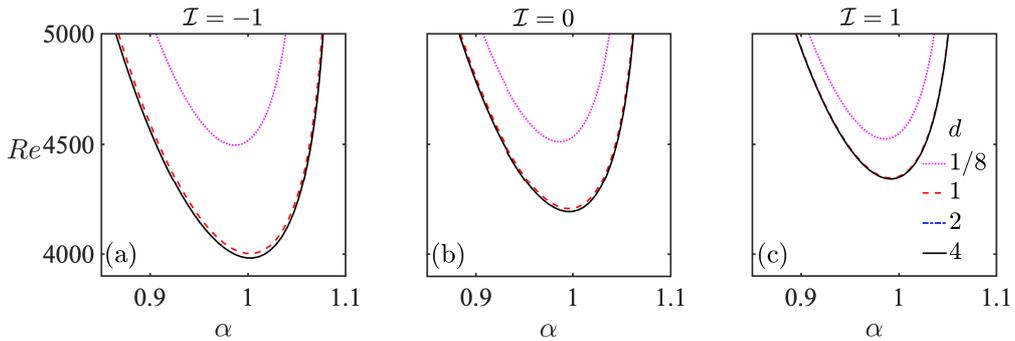}
    \caption{Neutral stability curves in the $(\alpha,Re)$--plane for different values of depth ratio with $\mathcal{I}=$ (a) $-1$, (b) $0$, and $1$.}
    \label{fig:neutral_inho}
\end{figure}
Figure~\ref{fig:neutral_inho} shows the neutral stability curves in the $(\alpha,Re)$-plane for three values of inhomogeneity factor $\mathcal{I} \in \lbrace -1,0,1 \rbrace$ and various values of the depth ratio $d$.
% At a given $\mathcal{I}$, as $d$ increases, the unstable regions become larger and the range of unstable wavenumbers increases.
Similar to the anisotropy, the destabilizing effect of the depth ratio continues to exist as the unstable zones expand as $d$ increases.
When $d \ge 1$, the neutral stability curves almost coincide at each $\mathcal{I}$, and thereby the associated critical Reynolds numbers are nearly insensitive to the depth ratio $d$. It is also seen from figure~\ref{fig:neutral_inho} that for each $d$, the unstable regions become larger, and the range of unstable wavenumbers increases as the inhomogeneity factor $\mathcal{I}$ decreases. Similar behaviour of stability boundaries are verified for other values of anisotropy parameter $\xi$.

\begin{table}[!htbp]
\centering
 \begin{tabular}{ |c||p{1.4 cm}|p{1.4 cm}|p{1.4 cm}|p{1.4 cm}|p{1.4 cm}|p{1.4 cm}|}
 \hline
 \multirow{2}{4em}{$\qquad d$}&\multicolumn{2}{|c|}{$\mathcal{I}=-1$}& \multicolumn{2}{|c|}{$\mathcal{I}=0$} &\multicolumn{2}{|c|}{$\mathcal{I}=1$}\\
 \cline{2-7}
         & $Re_c$& $\alpha_c$& $Re_c$& $\alpha_c$& $Re_c$ & $\alpha_c$\\
 \hline \hline
1/8 & 4469.39 & 0.9864 & 4511.71 & 0.9864 & 4524.85 & 0.9864\\
1   & 4001.39 & 1.0015 & 4205.73 & 0.9950 & 4348.26 & 0.9915\\
2   & 3981.37 & 1.0015 & 4192.15 & 0.9960 & 4344.09 & 0.9915\\
4   & 3981.19 & 1.0015 & 4191.89 & 0.9960 & 4344.15 & 0.9915\\
\hline
\end{tabular}
\caption{ Effect of inhomogeneity factor $\mathcal{I}$ and depth ratio $d$ on the critical parameters. %Reynolds number and critical wavenumber.
}
\label{table:neutral_I_d}
\end{table}
The critical Reynolds number $Re_c$ and critical wavenumber $\alpha_c$ are obtained for various $(\mathcal{I},d)$ and listed in Table~\ref{table:neutral_I_d}. The table indicates that $Re_c$ reduces by $0.938\%$ at $\mathcal{I}=-1$ compared to its value at $\mathcal{I}=0$ (homogeneous) for a small depth ratio $d=1/8$. The same trend is observed at higher $d=4$ (reduces $5.026\%$ for $\mathcal{I}=-1$ compared to $\mathcal{I}=0$), when thickness of each porous layer is fourth times to the thickness fluid layer. However, for larger inhomogeneity factor $\mathcal{I}=1$, there is an increase of $0.291\%$ and $3.632\%$ at $d=1/8$ and $d=4$, respectively. 

\begin{figure}[!htbp]
    \centering
    \includegraphics[scale=0.29]{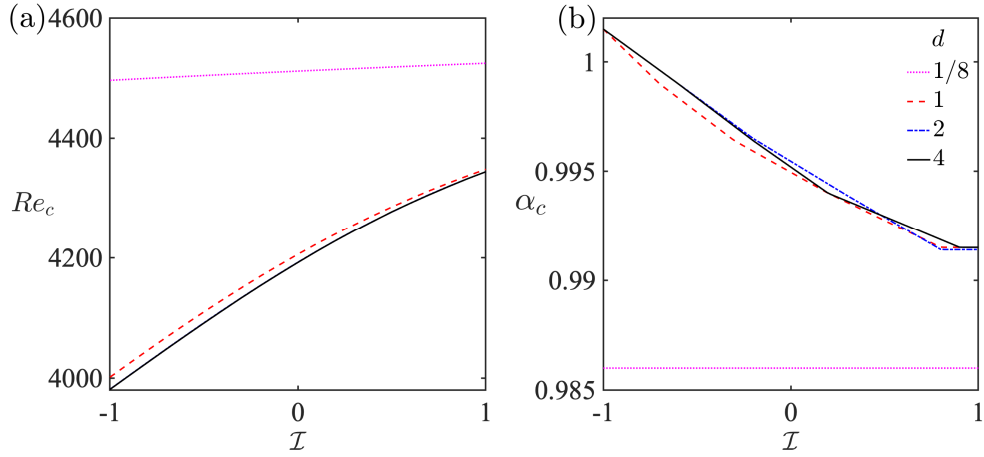}
    \caption{Variation of the (a) critical Reynolds number and (b) critical wavenumber with the inhomogeneity factor $\mathcal{I}$ for various depth ratio $d$.}
    \label{fig:critical_inho}
\end{figure}

In the end, the variation of critical parameters $Re_c$ and $\alpha_c$ with inhomogeneity factor $\mathcal{I}$ for various depth ratios $d$ are displayed in figures~\ref{fig:critical_inho}(a) and~\ref{fig:critical_inho}(b), respectively. $Re_c$ ($\alpha_c$) is a linear function of $\mathcal{I}$ with a very modest slope when $d=1/8$, i.e., for very thin porous walls. It is seen that the critical Reynolds number $Re_c$ (critical wavenumber $\alpha_c$) increases (decreases) as $\mathcal{I}$ increases for all depth ratios considered in the present study. 
The above-presented results are summarized as follows. A reduction in $\mathcal{I}$ translates to an increase in the overall permeability of the porous layers. Consequently, the flow rate increases, which reduces flow resistance, and hence flow becomes more unstable.
}

%-----------------------------------------------------------
\section{Conclusions and Outlook}
\label{sec:conclusion}
%-----------------------------------------------------------
Linear stability of a plane Poiseuille flow in a porous--fluid--porous configuration system with anisotropic and inhomogeneous permeability has been analyzed. The generalized Darcy's law has been used to model the dynamics in the porous layers and Navier--Stokes equation to model the dynamics of the open fluid-layer. The base velocity has been found analytically. The linear stability of the base flow to two-dimensional infinitesimal disturbances has been analyzed. The resulting Orr--Sommerfeld eigenvalue problem has been solved numerically using the Chebyshev spectral collocation method and the QZ--algorithm. Our preliminary computations have revealed that the primary instability is triggered by the fluid-layer mode for a range of parametric values considered. Furthermore, the stability results of anisotropic and inhomogeneous permeability variation have been compared with the isotropic and homogeneous permeability. It has been found that the directional and spatial variation of permeability, i.e., anisotropy and inhomogeneity, respectively, affect the flow instability significantly. By suitably manipulating the porous walls, one can modify and control the stability characteristics of the confined porous channel flow.

The critical Reynolds number $Re_c$ is higher (lower) with increasing (decreasing) the anisotropy parameter. Note that an increase in $\xi$ corresponds to either a relative increase in permeability in the wall-parallel direction ($k_x$) or a relative decrease in permeability in the wall-normal direction ($k_y$). Due to a rise in $\xi$, the flow resistance becomes sufficiently higher, delaying the onset of instability. On the other hand, a relative increase in wall-normal permeability, i.e., decreasing $\xi$ reduces the flow resistance, thereby advancing the onset of instability compared to a confined isotropic porous channel.
It has also been found that a decrease (increase) in inhomogeneity factor $\mathcal{I}$, decreases (increases) the critical Reynolds number $Re_c$ and destabilizes the flow. An inhomogeneous variation of permeability increases (decreases) the porous medium's overall permeability and reduces (increases) the flow resistance, destabilizing (stabilizing) effects on the flow. Furthermore, present study has revealed that the channel with thicker porous walls is more unstable than the thinner porous walls.

{ Apart from the anisotropic and inhomogeneous permeability, and depth ratio of the porous wall in controlling the instability of the flow in a confined porous-fluid-porous, as demonstrated in the present study, there is at least one instance wherein the magnetic field acts as a control parameter. Such effects are commonly studied under {\it magnetohydrodynamics}. Several studies have demonstrated 
% the effect of the Hartman number (the ratio of electromagnetic force to the viscous force) 
the stability of hydromagnetic steady flow between two parallel plates~\cite{MAKINDE2003Magneto,MAKINDE2007Temporal,MAKINDE2009temporalporous,makinde2002hydromagnetic}; those are either plane Poiseuille flow, generalized Couette--Poiseuille flow, or a flow through a permeable bed. The magnetic field intensity and a decrease in porous layer permeability while flowing through a porous bed has a stabilizing effect on the fluid flow. The inclusion of the effects of the magnetic field for the present case will be considered in the future.
}

The present investigation motivates the experimental study to examine the instability in a pressure–driven flow in confined porous channels. However, the nonlinear analysis generally gives more significant results and will be presented in the future. The present linear stability predictions can be utilized to avoid instability in practical applications such as manufacturing composite materials. Furthermore, the results from the present study can be used to manufacture and develop smaller types of equipment required for heat exchanges where instability is desirable.

% -----------------------------------------------------------------------
\appendix
% -----------------------------------------------------------------------
\section{Linear stability results for asymmetric velocity profiles}
\label{subsec:asymmetric_laminar}
% -----------------------------------------------------------------------
{

This appendix presents the linear stability results for asymmetric velocity profiles by
considering the upper porous layer as impermeable and the lower porous layer as permeable. 
\begin{figure}[!htbp]
    \centering
    \includegraphics[scale=0.3]{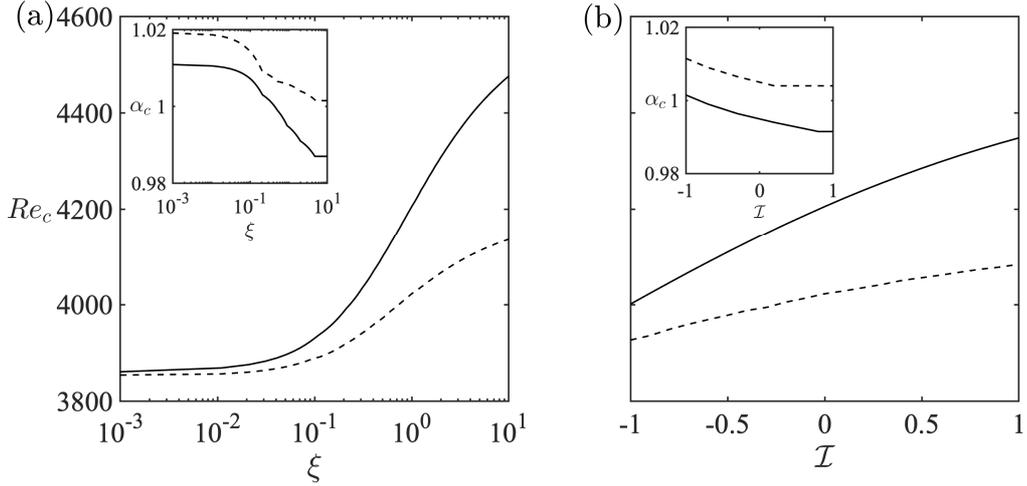}
    \caption{Comparison of (a) the critical Reynolds number $Re_c$ as a function of anisotropy parameter $\xi$ when $\mathcal{I}=0$ and (b) the critical Reynolds number $Re_c$ as a function of inhomogeneity factor $\mathcal{I}$ when $\xi=1$ in a channel with two porous walls (solid line, $\sigma=0.001,\; d=1$) and a channel with only one porous wall (dashed line, $\sigma_1=0.001,\; d_1=1$).}
    \label{fig:asymmetric}
\end{figure}
Figures~\ref{fig:asymmetric} (a) and~\ref{fig:asymmetric} (b) show the stability results of a channel with an impermeable upper wall and a permeable lower wall characterized by the parameters $\epsilon_1=0.3$, $\alpha_1^{BJ}=0.1$, $\sigma_1=0.001$ and $d_1=1$ (dashed line), and allow for a comparison presented in figures~\ref{fig:critical_anisotropy} and~\ref{fig:critical_inho} for a channel with two porous walls (solid line).
% It is seen from figure~\ref{fig:asymmetric}(a) that the critical Reynolds number $Re_c$ and the critical wavenumber $\alpha_c$ with only lower permeable homogeneous porous wall are smaller and higher than the critical Reynolds number $Re_c$ and the critical wavenumber $\alpha_c$ in a channel with two permeable homogeneous ($\mathcal{I}=0$) porous walls.
Compared to a channel with two permeable homogeneous ($\mathcal{I}=0$) porous walls, the critical Reynolds number $Re_c$ and the critical wavenumber $\alpha_c$ for a homogeneous lower permeable wall are less and higher, respectively, as shown in figure~\ref{fig:asymmetric}(a). Now, we set $\xi=1$ and find $Re_c$ and $\alpha_c$ as a function of inhomogeneity factor $\mathcal{I}$ for a channel with two permeable porous walls and a single lower porous wall. Although the critical Reynolds number $Re_c$ (critical wavenumber $\alpha_c$) grows (decreases) with inhomogeneity factor $\mathcal{I}$, the critical Reynolds number $Re_c$ (critical wavenumber $\alpha_c$) for a channel with a single porous wall is smaller (larger) than a channel with two porous walls for each $\mathcal{I}$. From the above, it is concluded that the anisotropic and inhomogeneous permeability has less effect on the critical Reynolds number and critical wavenumber in a channel with a single porous wall than in a channel with two porous walls.
}

 \bibliographystyle{elsarticle-num-names}
 \bibliography{refer}
\end{document}